\documentstyle[aps,preprint,epsfig,floats]{revtex}

\newcommand{\equ}[1]{(\protect\ref{#1})}
\newcommand{\bA}[1]{\hat{b}_#1}
\newcommand{\bC}[1]{\hat{b}_#1^\dagger}
\newcommand{\aA}[1]{\hat{a}_#1}
\newcommand{\aC}[1]{\hat{a}_#1^\dagger}
\newcommand{\state}[1]{\left| {#1} \right>}

\begin{document}

\draft 

\tightenlines
  
\title{Field theory for a reaction-diffusion model\\ 
  of quasispecies dynamics}

\author{Romualdo Pastor-Satorras$^1$ and Ricard V. Sol{\'e}$^{1,2,3}$}

\address{$^1$Complex Systems Research Group, Departament of Physics, FEN\\
Universitat Polit{\`e}cnica de Catalunya, 
Campus Nord B4, 08034 Barcelona, Spain\\
$^2$NASA-associated Astrobiology Institute, Carretera del Ajalvir Km
4, Madrid\\ 
$^3$Santa Fe Institute, 1399 Hyde Park Road, New Mexico 87501, USA}

\date{\today}
\maketitle

\begin{abstract}
  RNA viruses are known to replicate with extremely high mutation
  rates. These rates are actually close to the so-called error
  threshold. This threshold is in fact a critical point beyond which
  genetic information is lost through a second-order phase transition,
  which has been dubbed the ``error catastrophe.'' Here we explore
  this phenomenon using a field theory approximation to the spatially
  extended Swetina-Schuster quasispecies model [J. Swetina and P.
  Schuster, Biophys. Chem. {\bf 16}, 329 (1982)], a single-sharp-peak
  landscape. In analogy with standard absorbing-state phase
  transitions, we develop a reaction-diffusion model whose discrete
  rules mimic the Swetina-Schuster model. The field theory
  representation of the reaction-diffusion system is constructed.  The
  proposed field theory belongs to the same universality class than a
  conserved reaction-diffusion model previously proposed [F. van
  Wijland {\em et al.}, Physica A {\bf 251}, 179 (1998)]. From the
  field theory, we obtain the full set of exponents that characterize
  the critical behavior at the error threshold. Our results present
  the error catastrophe from a new point of view and suggest that
  spatial degrees of freedom can modify several mean field predictions
  previously considered, leading to the definition of characteristic
  exponents that could be experimentally measurable.
\end{abstract}

\pacs{PACS numbers: 87.10.+e, 02.50.-r, 64.60.Ak}

\section{Introduction}

RNA viruses offer a unique opportunity for exploring long term
evolution under controlled conditions due to their high mutation rates
\cite{preston96,domingo97}. Their evolutionary success is to a large
extent due to their small-sized genomes, but specially to their
enormous plasticity and adaptability to changing environments
\cite{domingo94}.  These viruses display the highest possible mutation
rates and, as a consequence, their populations, the so-called {\em
  molecular quasispecies} \cite{eigen89}, are extremely heterogeneous.
The quasispecies structure has numerous implications for the biology
of viruses. The most relevant of them is that mutant swarms are
reservoirs of variants with potentially useful phenotypes in the face
of environmental change. As extremely simplified entities at the
border of a life-like state, they are specially apt to mathematical
modeling \cite{alves98,nowak89}.

The replication of these molecules implies two basic reactions
\cite{eigen89,rowe94}: (i) error-free copy, when a (molecular) species
$I_i$ replicates by using available monomers $(A)$, i.e.:
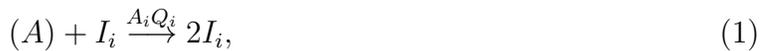
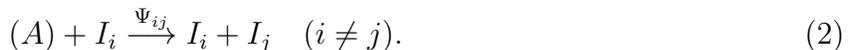
\begin{equation}
  (A) + I_i  \buildrel A_i Q_i \over  \longrightarrow 2 I_i,
\end{equation}
and (ii) mutation:
\begin{equation}
  (A) + I_i \buildrel  \Psi_{ij} \over  \longrightarrow I_i + I_j \quad (i\neq j).
\end{equation}
The parameters $A_i$ and $Q_i$ are the replication rate and the
quality factor, respectively. $Q_i \in [0,1]$ is a measure of the
correctness of the replication process, and it is maximum ($Q_i =1$)
if no mutations occur. $\Psi_{ij}$ are the mutation rates which can lead
to transitions between species $j \to i$.

The standard approach to the quasispecies dynamics (in the limit of
very large populations) is based on the continuous Eigen model
\cite{eigen89}.  Here a set of molecules which can replicate and
mutate is considered.  The basic equations are:
\begin{equation}
  {dx_i \over dt} = (A_iQ_i - D_i) x_i + \sum_{j \neq i} \Psi_{ij} x_j + \Phi_i,
  \label{eq:originaleigen}
\end{equation}
where $x_i$, with $i=1,2, ... , n$, accounts for the population size
of each species, $D_i$ stands for spontaneous degradation of molecules
(assumed to be linear), and $\Phi_i$ is an outflow term which takes into
account the removal of molecules from the system. If we introduce the
constraint of constant population size (CP), $\sum_i x_i = \mbox{\rm
  const.}$, the previous equations read:
\begin{equation}
{dx_i \over dt} = (A_iQ_i - D_i - \bar{E}) x_i + \sum_{j \neq i} \Psi_{ij} x_j,
\end{equation}
where the mean value of the so-called excess productivity,
$E_i=A_i-Q_i$, is given by $\bar{E} = \sum_i (A_i - D_i) x_i/\sum_j x_j$.

One of the most important results of Eigen's theory was the finding
that a phase transition takes place when mutation rates are tuned.
Specifically, let us assume for simplicity that each species is
composed by a string of elements $\{S_1, ..., S_{\nu}\}$ of size $\nu$
\cite{notegenom}. We consider in total a population of $N$ strings.  A
particular kind of sequences, composed by the elements $\{S_1^{(0)},
..., S_{\nu}^{(0)}\}$, represent the correct genomic sequences of the
species, and is called the {\em master sequence}.  Each time a string
is chosen to be replicated, it does so with some sequence-dependent
probability $r_i=P(\{ S_i \}$).  If replication occurs, each unit can
mutate with probability $\mu$, and reproduce exactly with probability
$1-\mu$.  Mutation introduces disorder into the system, and it can be
shown that a well-defined mapping exists between replication dynamics
and the two-dimensional Ising model in such a way that the temperature
is given by $T \approx - \vert \log (\mu/(1-\mu))\vert$
\cite{leuthausser86,leuthausser87,tarazona92}. For small $\mu$ the
replication system reaches a steady state in which the probability of
observing the master sequence is finite. On the other hand, for values
of $\mu$ larger than a given threshold, the probability of observing the
master sequence is vanishing small.

Eigen's theory predicts that genetic information becomes lost for
mutation rates higher than the critical rate $\mu_c$, due to a breakdown
of heredity and the lack of selection---the so-called ``error
catastrophe.''  It has been shown that this phenomenon indeed occurs
in RNA viruses, which replicate close to the error threshold 
\cite{domingo94}.  Actually, experimental data reveal that real viruses have
mutation rates $\mu \approx 1/\nu$, that is, inversely proportional to the size
of the genomic content, consistently with the prediction, and this has
led to the claim that increased mutation rates might be able to bring
virus populations into extinction. Such a strategy has been recently
shown to hold {\em in vitro} and is likely to be feasible {\em in 
vivo} \cite{loeb99}.

The presence of a critical mutation rate allows to interpret the error
catastrophe in the framework of standard {\em absorbing-state phase
  transitions} (APT) \cite{marro99,grinstein97}. APT are a class of
non-equilibrium transitions in which, by the variation of a control
parameter, the system crosses from an active phase with everlasting
activity, to an absorbing phase, in which the system remains trapped
forever, with no possibility to escape. In the framework of
species replication with mutation, the active phase is identified with
the low mutation regime, while the absorbing phase corresponds to the
high mutation regime.  Most APT are phase transitions of second order.
If we characterize the system by an appropriate order parameter $\psi$,
which in this case corresponds to the density of master sequences in
the system, by tuning the parameter $\mu$ we observe the typical
behavior
\begin{eqnarray*}
  \psi = 0, \qquad \qquad \quad&{\rm for}& \; \mu> \mu_c,\\
  \psi  \simeq (\mu_c - \mu)^\beta   , \qquad &{\rm for}& \;  \mu< \mu_c,
\end{eqnarray*}
close to the critical point $\mu_c$. The previous expression serves to
define the critical exponent $\beta$. The analogy with error catastrophe
is in this sense clear: for mutation rates larger than the error
threshold, the virus is inviable and it quickly dies. For small
mutation rates the virus is able to survive and reaches viable
populations whose size is an increasing function of $\mu_c - \mu$.
Further extending the analogy with APT, we can consider the spatial
and time dependence of the order parameter $\psi$, and define the
correlation function $g(r, t) = \left< \psi(r', t')\psi(r'+r, t'+t)
\right>$, where the bracket denote averages over different
realizations of the system. According to the dynamic scaling ansatz
\cite{cardy96}, we expect to observe close to the error threshold the
behavior 
\begin{equation}
  g(r, t) = r^{-(d+\eta)} F\left(\frac{r}{\xi}, \frac{t}{\xi^z}\right)
  \label{eq:correl1}
\end{equation}
which defines the correlation length $\xi$, related to the distance to
the $\mu_c$ by
\begin{equation}
  \xi \sim (\mu_c - \mu)^{-\nu_\perp}.
  \label{eq:correl2}
\end{equation}
Eqs.~\equ{eq:correl1} and \equ{eq:correl2} define the new critical
exponents $\eta$, $\nu_\perp$ and $z$, which determine the scaling of the
correlation function with respect to changes in the mutation rate
$\mu$.

Based in the previous analogy, in this paper we propose to study the
phenomenon of the error catastrophe from the point of view of an APT
by analyzing a reaction diffusion model with captures the essence of
the replication-plus-mutation mechanism of quasispecies dynamics. The
model allows the construction of an associated field theory,
representative of the same universality class, following a standard
technique outlined in the work of Doi, Cardy, and others
\cite{doi76a,doi76b,peliti85,lee95}.  The field theory developed here
is shown to correspond to a conserved reaction diffusion model
previously proposed by van Wijland {\em et al} \cite{wij98} (see also
Ref.~\cite{kree89}), in which the critical exponents were obtained
performing a one-loop renormalization group analysis.

An important aspect, seldom considered in previous studies, is the
effect of spatial degrees of freedom in quasispecies models. An exception 
is Adami's work on artificial life systems, in which a set of replicating 
bit strings of code spread on a two-dimensional lattice
\cite{Adami}. Under appropriate conditions, it was shown that the population 
spontaneously evolves to the error threshold, although no characterization 
of the model behavior at this critical point was performed. Besides, 
only a few experimental studies have recently reported the presence of
several patterns of virus distribution that cannot be explained in
terms of spatially-implicit quasispecies dynamics \cite{RNAexp,RNAexp2}. 

The interest in this problem is twofold: On the one hand, virus
populations show heterogeneity in space, thus introducing further
complexity in quasispecies dynamics and creating new opportunities to
viral evolution. On the other hand, it would be important to know if
spatially extended, mean-field models, are appropriate descriptions of
the real quasispecies dynamics in space.  Although RNA fitness
landscapes are known to be rugged \cite{burch99,burch00}, here we
consider the simplest, single-sharp-peak landscape \cite{swetina82}.
This model has been used as a null model of quasispecies populations
and a field theory of the reaction-diffusion rules can be developed.

The paper is structured as follows. In Sec. II we propose, in analogy
with the Swetina-Schuster model \cite{swetina82,sole99} a streamlined
reaction-diffusion model which captures the minimal elements in the
reproduction/mutation mechanism of quasispecies dynamics. Sec. III
reports a mean-field analysis, which allows to pinpoint the key
parameters of the model. In Sec. IV we construct the field theory
corresponding to the model. In analogy with the analysis performed by
van Wijland {\em et al.} and Kree {\em et al.} \cite{wij98,kree89}, we
obtain the relevant critical exponents.  Finally, we interpret our
results and put forward several experimental applications of them in
Sec. V.

\section{Reaction-diffusion model}

The key ingredient of our model consists in the assumption that one of
the sequences $B \equiv I_m$ has a high replication rate, while {\em all}
the others $A \equiv I_{j\neq m}$ have the same, lower replication rate. The
first sequence is called the master sequence and this approximation
defines the so-called Swetina-Schuster model \cite{swetina82}. Second,
by assuming that the sequences are long enough (consistently with real
RNA viruses, where $\nu>10^4$), backward mutations from $A$ to $B$ can
be neglected \cite{notecomment}.

In order to propose the reaction steps defining the model, we consider
a simplified version of the Swetina-Schuster model (see
Ref.~\cite{sole99} and references therein). In this model, a
population of $N$ strings of size $\nu$ evolves by a mechanism of
replication with errors. Each string is defined by a sequence $S_1,\ldots
S_\nu$, with $S_i = \{ 0,1 \}$. At each time step, we select a string and
replicate it, after removing another string chosen at random. The
replication takes place with probability $1$ for the master sequence,
defined by $S_i=1, \forall i$, and with probability $p<1$ for the rest. The
replication procedure replaces each element $S_i$ of the string for
$S'_i = S_i$ with probability $1-\mu$, and for $S'_i = (S_i+1)\; {\rm mod}
\; 2$ with probability $\mu$. This version of the model shows in plain
view the elementary steps of the error catastrophe: replication of the
master sequence at a certain rate, mutation of the master sequence,
and lack of backward mutation, for sufficiently large sequence length.
Another important ingredient in the model to be remarked is the
implicit constraint of constant number of sequences, realized in the
random deletion step, and which is usually implemented in the
quasispecies analytical models \cite{eigen89}. The model described in
Ref.~\cite{sole99} seems to display all the features of the error
catastrophe.

%Thus, on universality grounds, a reaction-diffusion
%version of this model is also expected to exhibit the same behavior
%and, most importantly, to belong to the same universality class.

Using this simple framework, we can translate the dynamics of the
Swain-Schuster model at a microscopic level in terms of reactions
among particles of type $B$ and $A$, corresponding to the master
sequence and the mutants. respectively. The simplified model in
Ref.~\cite{sole99} implicitly introduces interactions among sequences,
by means of the random deletion step. Thus, our reaction diffusion
model considers all the possible binary reactions between particles of
type $B$ and $A$, that are compatible with the outcome of the rules
used in \cite{sole99}. The set of reactions that we consider is:
\begin{mathletters}
  \label{eq:baremodel}
\begin{eqnarray}
  B + B & \stackrel{\mu}{\longrightarrow}& B + A 
  \label{eq:baremodel1}\\
  B + A & \stackrel{1-\mu}{\longrightarrow}& B + B
  \label{eq:baremodel3}\\
  A + B & \stackrel{\lambda}{\longrightarrow}& A + A
  \label{eq:baremodel4}
\end{eqnarray}
\end{mathletters}
The steps represent the replication/mutation of the first species,
coupled with the random deletion of the second species. Thus, the
reaction \equ{eq:baremodel1} implements the replication with mutation
of a master sequence $B$, which happens with an effective mutation
rate $\mu$, coupled to the random deletion of a sequence of type $B$;
the reaction \equ{eq:baremodel3} represent the exact replication of a
master sequence, at rate $1-\mu$, with the deletion of a $A$ sequence;
finally, the reaction \equ{eq:baremodel4} stands for the exact
replication of a sequence $A$, at rate $\lambda$, together with the deletion
of a master sequence.  All the remaining binary reactions with
replication/mutation plus random deletion do not alter the total
number of particles, and are thus not considered. The proposed set of
reactions mimic the conserved nature of the model imposed by the
random deletion of sequences in Ref.~\cite{sole99}, in a more nature
way than in the original quasispecies model,
Eq.~\equ{eq:originaleigen}, in which one had to impose an external
flow term $\Phi_i$ in order to ensure conservation.

The set of equations \equ{eq:baremodel} constitutes the core of our model.
Spatial effects are taken into account by allowing the different
particles to diffuse with respective diffusivities $D_A$ and $D_B$
\cite{notdiffusion}.

Given the interpretation of the different
particles, it is natural to consider $D_A=D_B$, that is, both master
sequence and mutants diffuse with the same speed. However, for the
sake of completeness, we will develop the formalism with $D_A\neq D_B$,
and make them equal only as a last step.

\section{Mean-field analysis}

In order to gain some preliminary intuition on the behavior of the
model, we analyze it by applying a standard mean-field analysis. Let
us denote by $\rho_B$ and $\rho_A$ the density of species $B$ and $A$,
respectively. Since the reactions \equ{eq:baremodel} conserve the number
of particles, the total density $\rho_B + \rho_A$ is constant in time:
\begin{equation}
  \rho_B + \rho_A = \rho.
  \label{eq:conserv}
\end{equation}
The classic (mean-field) equations for the densities  $\rho_B$ and $\rho_A$ are
readily found to be:
\begin{mathletters}
\begin{eqnarray}
  \frac{\partial \rho_B}{\partial t} & = &  (1- \lambda  - \mu) \rho_B \rho_A - \mu \rho_B^2, \\
  \frac{\partial \rho_A}{\partial t} & = & -(1- \lambda - \mu) \rho_B \rho_A + \mu \rho_B^2.
\end{eqnarray}
\end{mathletters}
Combining this with the conservation condition \equ{eq:conserv} we
obtain a single equation for the density $\rho_B$ of master sequences:
\begin{equation}
  \frac{\partial \rho_B}{\partial t} =  (1- \lambda - \mu) \rho \rho_B - (1- \lambda)  \rho_B^2.
  \label{eq:mean}
\end{equation}
This equation has two stable stationary states, depending on the value of
$\mu$:
\begin{eqnarray}
  \rho_B = 0, \qquad &{\rm for}& \; \mu> 1- \lambda,\\
  \rho_B = \frac{1- \lambda - \mu}{1-\lambda} \rho  \qquad &{\rm for}& \;  \mu< 1-\lambda.
\end{eqnarray}
At the mean-field level we observe the presence of a standard
absorbing-state phase transition at a critical point $\mu_c = 1- \lambda$. In
the subcritical regime, $\mu>\mu_c$, the order parameter (in this case the
density of master sequences) vanishes; in the supercritical region,
$\mu<\mu_c$, the order parameter has a power-law dependence on $\mu$:
\begin{equation}
  \rho_B \simeq   (\mu_c - \mu)^\beta,
\end{equation}
which defines the critical exponent $\beta$ in the mean-field
approximation, $\beta_{\rm MF} = 1$. As we will see in the next section,
the presence of fluctuations will change the value of $\beta$ at the
relevant, experimental, dimensions. A very interesting property of
this conserved RD model is that the critical point is independent of
the total particle density $\rho$, and is given as a function only of the
reproduction rate $\lambda$. This situation should be compared with the
conservative RD systems proposed so far, in which the total particle
density plays the role of the tuning parameter, and must be tuned to a
critical density $\rho_c$ in order for the system to display critical
behavior \cite{wij98,pv00}.

The mean-field solution also provides the expression for the probability
${\cal P}$ that there is at least one master sequence in the
steady-state regime \cite{notesurv}:
\begin{equation}
  {\cal P} = \Theta(\mu_c - \mu),
\end{equation}
where $\Theta$ is the Heaviside function.

\section{Field theory}

Our model consists in a set of particles of type $B$ and $A$ diffusing
in a hypercubic lattice of mesh size $h$ and size $L$, that interact
probabilistically according to the rules \equ{eq:baremodel} whenever
they meet at the same lattice site. The dynamics of the model is
defined through a master equation for the probability $P(\{n\} ,\{m\}, t)$
of having a particle configuration $\{n\}$, $\{m\}$ of particles $B$ and
$A$, respectively, at time $t$. The configuration $\{n\}=\{n_1, n_2, \ldots,
n_{L^d} \}$ represents the occupation number of each node in the
lattice. The master equation for $P$ is:
\begin{mathletters}
  \label{eq:master}
  \begin{eqnarray}
    \lefteqn{\frac{\partial}{\partial t} P(\{n\} ,\{m\}, t)  =} \nonumber \\
    && \frac{D_B}{h^2} \sum_{i,j} \left[(n_j+1) P(\ldots n_i-1, n_j+1, \ldots, \{m\}, t) 
      - n_i P(\{n\} ,\{m\}, t)\right]\\
    &+& \frac{D_A}{h^2} \sum_{i,j} \left[(m_j+1) P(\{n\}, \ldots m_i-1, m_j+1, \ldots,  t) 
      - m_i P(\{n\} ,\{m\}, t)\right] \\
    &+& \mu \sum_i \left[(n_i+1) n_i P(\ldots n_i+1, \ldots m_i-1, \ldots ,t) -
      n_i(n_i-1)P(\{n\} ,\{m\}, t) \right] \\
    &+& (1-\mu)  \sum_i \left[ (n_i-1)(m_i+1) P(\ldots,n_i-1, \ldots,m_i+1, \ldots,t) - n_i
      m_i P(\{n\} ,\{m\}, t)\right] \\
    &+& \lambda \sum_i \left[ (n_i+1)(m_i-1) P(\ldots,n_i+1, \ldots,m_i-1, \ldots,t) - n_i
      m_i P(\{n\} ,\{m\}, t)\right],
  \end{eqnarray}
\end{mathletters}
where $D_B$ and $D_A$ are the diffusion coefficients for $B$ and $A$
particles, $i$ is summed over all the lattice sites, and $j$ over the
nearest neighbors of the site $i$. The first two terms in the rhs of
Eq.~\equ{eq:master} implement diffusion through a random hopping of
particles between nearest neighbor sites.  The initial condition
$P(\{n\} ,\{m\}, 0)$ is given by a Poisson distribution, with an average
density per site equal for both types of particles.

The next step consists in recasting the master equation into a
``second quantized'' form, following the procedure described by Doi
\cite{doi76a,doi76b,peliti85,lee95}. We introduce two sets of
annihilation and creation operators at each lattice site, $\bA{i}$ and
$\bC{i}$ for $B$ particles, and $\aA{i}$ and $\aC{i}$ for $A$
particles, which fulfill the standard commutation rules
\begin{equation}
  \left[ \aA{i}, \aC{j} \right] = \left[ \bA{i}, \bC{j} \right] = \delta_{ij}.
\end{equation}
With this commutations rules, the operators have a bosonic character,
that is natural given the multiple occupancy of sites allowed in the
model. With the help of the vacuum state $\state{0}$, defined by
$\aA{i}\state{0} = \bA{i} \state{0} =0$, we construct an orthonormal
basis of states $\state{n,m}$, defined by
\begin{equation}
  \state{n,m} = \prod_i (\bC{i})^{n_i} (\aC{i})^{m_i} \state{0},
\end{equation}
and work in the Fock space spanned by this basis. In terms of this
Fock space, the state of the system at time $t$ is represented by the
vector state $\state{P(t)}$, defined as
\begin{equation}
  \state{P(t)} = \sum_{\left\{n\right\},\left\{m\right\}} P(\{n\} ,\{m\},
  t)\state{n,m}. 
\end{equation}
In terms of this vector state, the master equation Eq.~\equ{eq:master}
can be rewritten as a Schr{\"o}dinger equation in imaginary time
\begin{equation}
  \label{eq:schro}
  \frac{\partial}{\partial t} \state{P(t)} = -\hat{H} \state{P(t)},
\end{equation}
with a Hamiltonian, or time-evolution operator, $\hat{H}$ defined by:
\begin{eqnarray}
  \label{eq:hamil}
  \hat{H} & = & \sum_{\left< i j \right>} \left[ \frac{D_A}{h^2} (\aC{i} -
    \aC{j})(\aA{i} - \aA{j}) + \frac{D_B}{h^2} (\bC{i} -\bC{j})(\bA{i} -
    \bA{j}) \right]  \\
  & + & \mu \sum_i (\bC{i} - \aC{i}) \bC{i} \bA{i}^2 + (1-\mu) \sum_i (\aC{i} - \bC{i})
  \bC{i} \bA{i} \aA{i} + \lambda  \sum_i
  (\bC{i} - \aC{i}) \aC{i} \bA{i} \aA{i} .
\end{eqnarray}
Eq.~\equ{eq:schro} can be formally solved in terms of the operator
$\hat{H}$ yielding
\begin{equation}
  \state{P(t)} = \exp ( -  \hat{H} t )
  \state{P(0)}.
  \label{eq:solutionop}
\end{equation}
From this solution, it is possible to derive all the statistical
properties of the RD system, applying a projection technique
\cite{doi76a,doi76b,peliti85,lee95}. For practical purposes, it is
convenient to map this second-quantized form into a field theory,
using a coherent state representation. Performing a time-slicing of
the evolution operator in Eq.~\equ{eq:solutionop}, via the Trotter
formula, we can express the vector state $\state{P(t)}$ as a path
integral, weighted with the exponential of an action S, over a set of
classical fields $a^*$, $a$, $b^*$, and $b$, which are related with
the two types of particles. After taking the continuum limit ($h \to
0$), the vector state can be written as the path integral over space
and time dependent fields
\begin{equation}
  \state{P(t)} =  \int {\cal D}  a \; {\cal D}a^* \;  {\cal D} b \;
  {\cal D} b^* \; \exp (- S[ a, a^*, b,  b^*]) \state{P(0)},
\end{equation}
where the action $S$ has the form \cite{noteaction}
\begin{eqnarray*}
  S[ a, a^*, b,  b^*] & = & \int d^d x \int d t \left\{  a^* [ \partial_t -D_A \nabla^2 ] a +
    b^* [ \partial_t -D_B \nabla^2 ] b  \right. \\
  & + & \left. \mu (b^* - a^*) b^* b^2  + (1-\mu)
    (a^* - b^*) b^* a b + \lambda  (b^* - a^*) a^* a b \right\}.
\end{eqnarray*}
Within this formalism, we can compute the average value of any
observable  $F(\{n\} ,\{m\})$ performing the path integral
\begin{equation}
  \left<F(t)\right> = {\cal C}  \int  {\cal D} a  \; {\cal D} a^* \;
  {\cal D} b \;  {\cal D} b^* \; F(a, b) \exp (- S[ a, a^*, b,  b^*]),
\end{equation}
where ${\cal C}$ is an appropriate normalization constant. 

The final step in the derivation of the field theory consists in
performing the shift
\begin{equation}
  a^* = 1 + \bar{a}, \qquad  b^* = 1+ \bar{b},
\end{equation}
and the change of variables
\begin{equation}
  \begin{array}{ll}
    \psi  = b,                       & \phi  = a + b - \rho, \\
    \bar{\psi } = \bar{b} - \bar{a}, \qquad& \bar{\phi} = \bar{a}.
    \end{array}
\end{equation}
The final action describing the RD system is
\begin{eqnarray}
   \lefteqn{S[ \psi , \bar{\psi}, \phi ,  \bar{\phi}] = } \nonumber \\
   && \int d^d x \int d t
   \left\{ \bar{\psi} [ \partial_t \psi  -D_B \nabla^2 \psi  - r \psi  + g_1 \psi^2  - g_2 \psi \phi ]
    +  \bar{\phi} [ \partial_t \phi -D_A \nabla^2 \phi   + \gamma  \nabla^2 \psi ] \right. 
  \label{eq:ft}  \\
    &+& \left. \bar{\psi}^2 [ -g_3 \psi  -v_1 \psi \phi  + v_2 \psi^2 ] +  \bar{\psi} \bar{\phi} [
    -g_4 \psi  - v_3 \psi \phi + v_4 \psi^2 ] \right\},  \nonumber
\end{eqnarray}
where we have defined the coupling constants
\begin{eqnarray*}
  r &=& g_4 = (1-\lambda-\mu)\rho, \\
  g_1 & = & v_4 = 1-\lambda,\\
  g_2 & = & v_3 = 1-\lambda-\mu,\\
  g_3 & = & \rho v_1 = (1-\mu)\rho,\\
  v_2 & = & 1,\\
  \gamma & = & D_A - D_B.
\end{eqnarray*}
The coupling constant $r$, $\gamma$, $g_i$, and $v_i$ are coarse-grained
versions of the microscopic reaction rates.  Since we are only
interested in the behavior of the system close to a critical point,
however, the actual value of these parameters is
irrelevant~\cite{wij98,kree89}.

A na\"{\i}ve power counting shows that the critical dimension of this
field theory is $d_c=4$, and that the coupling constants $v_i$ are
irrelevant, and can be in principle discarded. 
With this final form, it is easy to recognize that the action
\equ{eq:ft} represents the same field theory analyzed by van Wijland
{\em et al.} for the conserved reaction diffusion model
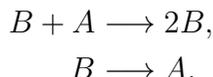
\begin{eqnarray*}
  B + A &\longrightarrow& 2B,\\
  B &\longrightarrow& A.
\end{eqnarray*}
In their study, the authors worked out the renormalization group
analysis for this system, for both cases: $D_A< D_B$ and $D_A=D_B$,
providing the critical exponents up to a one-loop expansion.
The case $D_A=D_B$, which also corresponds to the
universality class of a model of population dynamics with pollution
described by Kree {\em et al.} \cite{kree89}, is the relevant one in
the problem under consideration. Quoting the results of
Refs.~\cite{wij98,kree89}, we have the critical exponents:
\begin{mathletters}
\label{eq:exponents}
\begin{eqnarray}
  \beta = 1 - \frac{\varepsilon}{32} \\
  \nu_\perp  = \frac{1}{2 - \varepsilon /2} \\
  \eta = - \frac{\varepsilon}{8}\\
  z = 2,
\end{eqnarray}
\end{mathletters}
where $\varepsilon = 4-d$ gives the dimensionality of the system. The results
for $\nu_\perp$ and $z$ are exact, derived field-theoretically by analysing
the symmetries of the action \equ{eq:ft}, and are thus valid for all
dimensions. The values of $\beta$ and $\eta$, on the other hand, and
expansions around $\varepsilon=0$, and thus they are expected to hold only for
small values of $\varepsilon$.

In view of the results \equ{eq:exponents}, the relevant exponents at
the physical dimension $d=3$ are
\begin{equation}
  \begin{array}{ll}
    \nu_\perp  = 2/3 & \qquad z = 2, \\
    \beta \simeq 0.969 & \qquad \eta \simeq - 0.125.
  \end{array}
\end{equation}
The values of $\nu_\perp$ and $ z$ are exact, while those for $\beta$ and $\eta$
represent an approximation given by the replacement of the small
parameter $\epsilon$ for $1$. In dimension $d=2$ or less, the exponents
\equ{eq:exponents} have to be taken with a grain of salt, due to the
terms $v_i$ in Eq.~\equ{eq:ft}, which might become relevant
at low dimensions \cite{wij98}.

\section{Discussion}

The dynamical theory of molecular evolution developed by Eigen and
Schuster reveals the presence of an intrinsic, sharp limit to
molecular information carriers.  This threshold is a generic feature
of replicator systems involving reproduction and mutation.  The
Eigen-Schuster theory predicts that, under the effect of evolutionary
pressures selecting for high variability, such replicators will evolve
towards the error threshold. This is the case of RNA viruses and
experimental evidence clearly supports this theoretical prediction.

Previous theoretical models have analysed the stochastic dynamics of
quasispecies under different approaches. But all of them considered
spatially-implicit models (mean-field-like), paying no attention to
local effects derived from incomplete mixing. Here we have explored
this problem using the simplest quasispecies model, described by a
single-sharp-peak replication landscape. The aim of our study was to
see how the statistical behavior of a spatially extended molecular
replication system would differ from the mean-field predictions.

We have considered a simplified reaction-diffusion model where two
types of ``particles'' (the master sequence and the mutant sequences)
diffuse, replicate, and mutate on a given spatial domain. Applying the
standard approach to absorbing-state phase transitions, a field theory
has been developed and it has been shown to be the same reported by
Wijland {\em et al.} \cite{wij98} and Kree {\em et al.} \cite{kree89}.

The main message from our study is that relevant differences between
mean-field models and real dynamics are expected to be observed even
in the simplest scenario considered here.  Larger deviations should be
expected in more realistic models incorporating a better description
of molecular replicators and their dynamics. In particular, the
sharpness of the transition (as defined by the $\beta$ exponent) is not
very different at different dimensions. This suggests that no
measurable differences should be expected to be observed in
experimental systems. The correlation exponent $\nu_\perp$, however, does
change appreciably, from $\nu_\perp(d=1) = 2$ to $\nu_\perp(d=3) = 2/3$ thus leading
a faster decorrelation at realistic dimensions. Such an increase in
$\nu_\perp$ will enhance the coexistence of different strains (quasispecies)
in a given spatial domain \cite{SoleBascompte} and thus the
probabilities of success for the virus.

Several caveats of this approach are worth mentioning.  It is known
that real RNA viruses have actually multipeaked landscapes
\cite{burch99,burch00}.  We will expand our analysis to such situation
in a future work (although the associated field theory, constructed in
terms of several particles of type $B_i$, each one representing a
different viable master sequence, is expected to be much harder to
develop and analyze). However, in many situations the quasispecies are
observed to be confined in a fitness peak, so that our previous
analysis essentially holds \cite{burch99,burch00}.  Also, the real
RNA virus dynamics takes place through a virus-cell interaction not
considered here, while several previous theoretical models used in
order to understand well-defined experimental results (where a cell
population was present) have been shown to be successful in providing
a full understanding of the evolutionary dynamics of RNA populations
\cite{Tsimring,SoleJTB}.

\acknowledgments
\section*{}

The authors thank Santiago Elena, Peter Stadler, and Alessandro
Vespignani for valuable comments and suggestions.  This work has been
supported by a grant PB97-0693 and by the Santa Fe Institute (RVS).


\begin{references}
\bibitem{notegenom}
The string can be visualized as idealized genomic sequence defining the
  species, with a length of $\nu$ genes.
\bibitem{notecomment} The numerical analysis of the Swetina-Schuster
  model shows that this approximation works very well for $\nu>50$
  \cite{swetina82}. Sequences larger than this value provide a sharp
  phase transition. Since the smallest RNA molecules have a size of $\nu
  \approx 80$ (not to mention RNA viruses, whose genomes have sizes $\nu >
  10^3$) the current approximation seems well justified.
\bibitem{notesurv}
Note that this probability is not related with the surviving probability in
  spreading experiments usually defined in absorbing-state phase transitions
  \cite{marro99}.
\bibitem{noteaction}
Since we are only interested in the properties of the stationary state of the
  system, we have omitted in the expression of the action boundary terms that
  are relevant for the study of the relaxational dynamics; see for example
  Ref.~\cite{wij98}.
\end{references}
\end{document}